\documentclass[12pt]{article}
\usepackage{graphicx}
\usepackage{epsfig}
\usepackage{amsfonts}
\usepackage[lofdepth,lotdepth]{subfig}
\usepackage{url}
\usepackage{setspace}
\usepackage{multirow}
\usepackage{array}

\begin{document}

\title{Primitive Nonclassical Structures of the $N$-qubit Pauli Group}
\date{October 9, 2013}
\author{Mordecai Waegell \footnote{caiw@wpi.edu} \\ \it{ Physics Department, Worcester Polytechnic Institute} \\ {\it Worcester, MA, USA}}
\maketitle

\abstract{Several types of nonclassical structures within the $N$-qubit Pauli group that can be seen as fundamental resources for quantum information processing are presented and discussed.  Identity Products (IDs), structures fundamentally related to entanglement, are defined and explored.  The Kochen-Specker theorem is proved by particular sets of IDs that we call KS sets.  We also present a new theorem that we call the $N$-qubit Kochen-Specker theorem, which is proved by particular sets of IDs that we call NKS sets, and whose utility is that it leads to efficient constructions for KS sets.  We define the criticality, or irreducibility, of these structures, and its connection to entanglement.  All representative critical IDs for up to $N=4$ qubits are presented, and numerous families of critical IDs for arbitrarily large values of $N$ are discussed.  The critical IDs for a given $N$ are a finite set of geometric objects that appear to fully characterize the nonclassicality of the $N$-qubit Pauli group.   Methods for constructing critical KS sets and NKS sets from IDs are given, and experimental tests of entanglement, contextuality, and nonlocality are discussed.  Possible applications and connections to other work are also discussed}

\subsection{Introduction}

One of the central objectives in contemporary quantum information processing research is the discovery of static quantum resources \cite{NielsenChuang}, which is to say, nonclassical structures in Hilbert spaces.  In this paper we define and generalize several types of nonclassical structures within the $N$-qubit Pauli group.  The most important and fundamental of these are what we call Identity Products (IDs, for short), which are integrally related to $N$-qubit entanglement.  We also obtain structures that demonstrate quantum contextuality and nonlocality by constructing particular sets of these IDs.  \cite{WaegellThesis}.  In each case we also define conditions for the minimality or irreducibility of each structure, to ensure that the structures we obtain are the fundamental elements of their kind.  We call these irreducible nonclassical structures {\it critical}, and we will show that criticality is really a statement about the entanglement properties of these structures, which are the main focus of this paper.

Within the $N$-qubit Pauli group, the complete family of these critical nonclassical structures {\it is} the complete set of primitive elemental resources for quantum information processing.  Numerous applications of these resources will be discussed in later sections of this paper.

To begin, an ID is any set of mutually commuting $N$-qubit Pauli observables whose combined product is $\pm I$ (the identity in the product space of all $N$ qubits).  We will call an ID critical if it cannot be factored into the product {\it or} direct product of two or more smaller IDs.

Because a critical ID cannot be factored into the direct product of smaller IDs, it follows that the joint eigenbasis of the set of commuting $N$-qubit observables that form the ID must be maximally entangled (i.e. all $N$ qubits are inseparably entangled in those eigenstates).  This is an example of a general alternate definition of entanglement, in which we speak of the entanglement of a set of mutually commuting multipartite observables, rather than the entanglement of a state.  This is the most important part of what we mean when we say that an ID is critical.  A detailed analysis of the relationship between commutativity and entanglement is given in the appendix.  An ID has the additional feature that the product of these observables is fixed to be either $\pm I$, which means that the product of the measured eigenvalues for such a set of compatible observables is state-independent.

We will depict an ID composed of $M$ observables from the $N$-qubit Pauli group using the compact symbol ID$M^N$.  For critical IDs, the maximum value of $M$ is $N+1$, and in this case the ID is said to be complete, in that it fully defines a basis of maximally entangled rank-1 projectors.  Furthermore, when $M=N+1$, the complete stabilizer group for the entangled eigenbasis is generated by taking products of the observables in the critical ID.  This stabilizer group is itself always an ID, though it can only be critical for $N=2$, and can otherwise be factored into the product of smaller IDs.  The $N$-qubit Graph States \cite{BriegelGraphStates} are also associated with these stabilizer groups, and so each Graph State corresponds to some set of critical IDs.

There also exist critical IDs for $N\geq4$ for which $M<N+1$, and in this case we say the ID is incomplete, in that it does not fully constrain the eigenbasis of maximally entangled rank-$2^{N-M+1}$ projectors, each of which defines a subspace with internal degrees of freedom.  Nevertheless, the criticality of the ID still guarantees that any state within one of these subspaces is maximally entangled.  Critical IDs like this necessarily belong to more than one distinct stabilizer group, and thus form the intersection of such groups.

From an interpretational viewpoint, the set of representative critical IDs for a given number of qubits $N$ can be seen as a finite set of geometric modes in the space $\mathbb{C}^{2^N}$, in much the same way that the set of regular polytopes in $\mathbb{R}^d$ is finite (for $d>2$).  Furthermore, each critical ID of a given $N$ belongs to one or more stabilizer groups, and within these abelian subgroups, the ID can be seen as a sort of `prime' structure, in the sense that it cannot be factored into smaller structures in smaller stabilizer groups.

We have enumerated all representative IDs for up to $N=5$ using computational searches, and many examples for up to $N=16$.  We have further identified numerous symmetric families of critical IDs that generalize to arbitrarily large values of $N$ \cite{WaegellThesis, WA_Nqubits, WA_4qubits}.

Sets of IDs can be used to furnish proofs of the Kochen-Specker theorem \cite{Bell2, KS} that are based only on the observables of the $N$-qubit Pauli group \cite{WaegellThesis, WA_Nqubits, WA_4qubits, Mermin_SquareStar, WA_3qubits}, and as an intermediate step, to furnish proofs of a new theorem we call the $N${\it -qubit Kochen-Specker theorem} (NKS theorem), for reasons that will be explained below.  For brevity, we will refer to sets of IDs that prove the prior theorem as KS sets throughout this paper, and sets of IDs that prove the latter theorem as NKS sets.  These KS sets should not be confused with sets of projectors that go by the same name elsewhere in the literature.  \footnote{In \cite{WaegellThesis}, the $N$-qubit Kochen-Specker Theorem was instead given the name `The Strong Kochen-Specker Theorem,' which we have subsequently learned has already been used differently by other authors, and so we have adopted this new name instead.  We also referred to NKS sets as `Kernels,' a name we have dropped to avoid confusion with other uses of this word.}

The Kochen-Specker theorem is proved by a set of IDs $\{ID_j\}$ that contain $N$-qubit Pauli observables $\{O_i\}$, such that it is impossible to assign simultaneous noncontextual truth values $v(O_i) = \pm 1$ to all of the observables such that the product of the $v(O_j)$ in every ID is equal to the sign of that ID (i.e. the sign of the $N$-qubit identity obtained by taking the product of all $M$ observables in the ID$M^N$).  This impossibility occurs for a set of IDs iff the number of negative IDs in the set is odd, while each $N$-qubit Pauli observable appears in an even number of the IDs.  This follows because the overall product of all IDs in the set is negative, but in a noncontextual hidden variable theory, the product will contain only even powers of $v(O_i)$, making it positive.  This defines a KS set of IDs.

The $N$-qubit Kochen-Specker theorem is proved by a set of IDs $\{ID_j\}$ that contain $N$-qubit Pauli observables $\{O_i\}$, which are in turn composed of direct products of single-qubit Pauli observables $\{Z_q, X_q, Y_q\}$ and the single-qubit identity $I_q$ ($q=1,\ldots,N$), such that it is impossible to assign simultaneous noncontextual truth values $v(Z_q) = \pm 1$, $v(X_q) = \pm 1$, $v(Y_q) = \pm 1$, $v(I_q) = 1$ to all of the single-qubit observables such that the product of the $v(O_j)$ in every ID is equal to the sign of that ID.  This impossibility occurs for a set of IDs iff the number of negative IDs in the set is odd, while for each $q$, each single-qubit Pauli observable appears an even number of times throughout the set of IDs.  This follows because the overall product of all IDs in the set is negative, but in a noncontextual hidden variable theory that requires independent realism for every qubit, the product will contain only even powers of $v(Z_q)$, $v(X_q)$, and $v(Y_q)$, making it positive.  This defines an NKS set of IDs.

We will say that a KS (NKS) set is critical if it is impossible to remove any subset of IDs and/or entire qubits such that a smaller KS (NKS) set remains.  This latter requirement is important because a critical KS (NKS) set need not be composed of IDs that are critical in the sense we have defined for IDs.  Criticality of a KS (NKS) set only requires that all of the qubits be interconnected by a network of entanglement present across the IDs of the set, even if the individual IDs are not critical over all of the qubits in the KS (NKS) set.

The $N$-qubit Kochen-Specker theorem rules out theories for which each qubit must always be assigned independent truth values, which is to say, theories that deny the possibility of entanglement.  A critical NKS set cannot also be a complete KS set, and so alone it does not rule out the existence of noncontextual hidden variable theories that allow entanglement.  The usual Kochen-Specker theorem then rules out even these entanglement-friendly noncontextual hidden variable theories, but the critical KS set always requires additional IDs (i.e. additional measurement contexts).

Because we do not know of any counterexamples among the many KS sets we have discovered, we conjecture that every KS set contains a critical NKS set as a subset.  Furthermore, there are many ways to use a critical NKS set to generate critical KS sets.  If the conjecture holds, it is interesting to note that one cannot prove the KS theorem using $N$-qubit Pauli observables without also, and in some sense first, proving the NKS theorem.

The methods we develop here for examining quantum nonclassicality are somewhat divergent from other lines of research, but the real foundation for this body of work was laid by N. David Mermin \cite{Mermin_SquareStar} in his discussion of KS sets, and by Asher Peres and David DiVincenzo \cite{Peres_DiVinc} who discovered a connection between a 5-qubit quantum error correcting code, a critical single-ID NKS set, and ultimately a KS set, and even touched on the issue of projectors of ranks greater than unity.  We hope that the reader will find this new perspective on the subject to be stimulating.

The remainder of this paper is organized as follows.  We review the structure of critical IDs, and present all representative types for up to $N = 4$ qubits.  We also present several interesting examples of critical IDs for larger values of $N$.  The relationship between IDs, entanglement, stabilizer groups, and graph states is explored in more detail.  We then present methods for using critical IDs to construct NKS sets and KS sets, and discuss using these for experimental tests of entanglement, contextuality, and nonlocality.  Finally, we will discuss other possible applications for the nonclassical structures that are presented here.

For convenience, some of the larger tables and figures have been held for the appendix, though they are referred to throughout the text.  Furthermore, several of the figures are presented in color for additional clarity, but this may be lost in print versions of this paper.

\subsection{Identity Products (IDs)}

An ID$M^N$ is a set of $M$ mutually commuting observables from the $N$-qubit Pauli Group whose combined product is $\pm I$ (the identity in the space of all $N$ qubits).  We will depict IDs as tables in which it is understood that each row is a different observable in the set, and each column is a different qubit in the set, with the implied tensor product symbols between columns omitted for compactness.  Consider the ID$3^2$ of Table \ref{q2ID} as an example.  This ID is composed of 3 mutually commuting observables from the 2-qubit Pauli group, and the product of the 3 observables is $-I$, the identity in the product Hilbert space of both qubits.
\begin{table}[h!]
\begin{center} {
\begin{tabular}{cc}
$Z$ & $Z$\\
$X$ & $X$\\
$Y$ & $Y$\\
\end{tabular}}
\end{center}
\caption[]{}
\label{q2ID}
\end{table}

Each ID is composed of the tensor product of multiple ordered sets of single-qubit Pauli observables $\{Z,X,Y\}$ and the single-qubit identity $I$, which we call Single Qubit Products (SQPs).  The SQPs are the columns of the ID, and so each one is assigned to a particular qubit.  In order to belong to an ID, the SQP must fall into one of two classes:  1) Even SQPs are those in which each single-qubit Pauli observable appears an even number of times, and which have ordered product $\pm I$.   2) Odd SQPs are those in which each single-qubit Pauli observable appears an odd number of times, and which have ordered product $\pm i I$.  Because the product of the observables in an ID is real, the number of Odd SQPs in an ID is always even.  While the individual SQPs are ordered sets, a complete ID is no longer ordered because the observables all mutually commute - it is the relative order of the SQPs for each qubit that is important.

 This is because it is the relative order of the individual SQPs that ensures that the observables mutually commute.  The observables $\{Z,X,Y\}$ mutually anticommute, and so any two $N$-qubit Pauli observables that commute must also anticommute on an even number of their qubits.  The observables and qubits of a critical ID are inseparably linked together by these anticommutations (see Figures \ref{SQPgraphs}, \ref{IDGraphs1}, and \ref{IDGraphs2}).

Therefore a critical ID will never contain what we call Trivial SQPs, which are Even SQPs that contain either all $I$s, or some mix of $I$s and any one other Pauli observable $\{Z,X,Y\}$.  Trivial SQPs contain no anticommutation links and have ordered product $+I$, and so these qubits can always be discarded from an ID to leave a smaller ID (which makes the original noncritical by definition).  Therefore the only Even SQPs that appear in critical IDs are those containing at least two different Pauli observables.

In order to construct representative IDs from SQPs, we need not consider every order of the Pauli observables within each SQP - they mutually anticommute regardless of order or orientation - and so we only need to consider SQPs in which $\{Z,X,Y\}$ appear in that canonical order, with only their relative number and placement, and the placement of $I$s, making the SQPs distinct.  So finally, we define a {\it representative ID} to represent the complete class of IDs obtained from the canonical ID by taking all possible permutations of qubit order, and all possible rotations and reflections of each qubit's coordinate system.  All of the IDs in such a class have isomorphic anticommutation structure, whereas IDs belonging to different classes have nonisomorphic anticommutation structure.  We characterize this structure in terms of color-hypergraphs in Section \ref{IDGraphs}.

It should not be difficult to see that Table \ref{q2ID} is the only type of representative critical ID that exists for 2 qubits.  Similarly, it is fairly straightforward to see that Table \ref{q3IDs} contains the only 2 representative critical IDs that exist for 3 qubits.  Here we introduce the final detail in our symbol for IDs, a subscript indicating the number of Odd SQPs in the ID.  This detail is important for constructing NKS sets, as we will show in Section \ref{NKS}.
\begin{table}[h!]
\centering
\qquad
\subfloat[][ID$4^3_0$]{
\begin{tabular}{ccc}
$Z$ & $Z$ & $Z$ \\
$Z$ & $X$ & $X$    \\
$X$ & $Z$ & $X$   \\
$X$ & $X$ & $Z$    \\
\end{tabular}\label{q3ID1}}
\qquad
\subfloat[][ID$4^3_2$]{
\begin{tabular}{ccccccc}
$Z$ & $Z$ & $I$ \\
$Z$ & $I$ & $Z$    \\
$X$ & $X$ & $X$   \\
$X$ & $Y$ & $Y$    \\
\end{tabular}\label{q3ID2}}
\qquad
\caption[]{The two critical IDs of the 3-qubit Pauli group.  The ID$4^3_0$ of \subref{q3ID1} is the smallest example of an NKS set with only one ID.} \label{q3IDs}
\end{table}

Allowing for permutations of the qubit order and local transformations of the SQPs, both IDs in Table \ref{q3IDs} belong to common stabilizer groups, reflecting the fact that there is only 1 distinct type of 3-qubit entangled state generated by IDs of the 3-qubit Pauli group, namely the GHZ state \cite{GHZ}.

For 4 qubits, the situation becomes more subtle.  There are 9 representative critical IDs within the 4-qubit Pauli group.  All of the 8 representative IDs shown in Table \ref{q4IDs_8} belong to common stabilizer groups (allowing for permutations of the qubit order and local transformations of the SQPs), and these are the stabilizers of the 4-qubit Cluster states \cite{briegel2001persistent}.  The last representative ID, shown in Table \ref{q4ID_1}, is unique to the stabilizer groups of the 4-qubit GHZ states.  Clearly each class of entanglement within the $N$-qubit Pauli group corresponds to a particular set of critical IDs - a point we plan to explore in the future.  For the general class of $N$-qubit Graph states, each connected graph on $N$ vertices belongs to a particular class of entanglement in just the same way, as shown in the tables (all of the connected graphs on 2 and 3 qubits of course correspond to the 2-qubit Bell states and 3-qubit GHZ states respectively).  The complete enumeration of the 238 representative IDs for $N=5$, along with many other examples, may be found at our website \cite{MainWebsite}.  For details about demonstrations of nonclassicality using Graph States, see \cite{YuOhGraphStates, cabello2013exclusivity}.
\begin{table}[h!]
\centering
\qquad
\subfloat[][]{
\begin{tabular}{cccc}
$Z$ & $Z$ & $Z$ & $Z$ \\
$X$ & $X$ & $Z$ & $Z$ \\
$Y$ & $I$ & $X$ & $I$ \\
$I$ & $Y$ & $I$ & $X$ \\
$I$ & $I$ & $X$ & $X$ \\
\end{tabular}\label{q4ID_1}}
\qquad\qquad\qquad\qquad\qquad\qquad\qquad\qquad\qquad\qquad
\subfloat[][]{
\includegraphics[width=.75in]{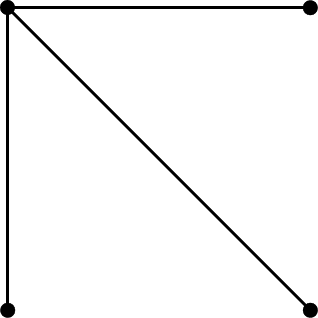}
\label{g4_2}}
\qquad
\subfloat[][]{
\includegraphics[width=.75in]{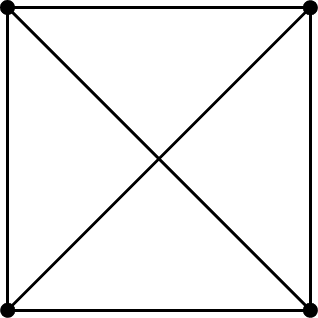}
\label{g4_6}}
\qquad
\caption{This lone critical ID$5^4_2$ belongs to the stabilizer groups of the 4-qubit GHZ states, as do these two 4-vertex graphs.}
\end{table}

The issue of checking an arbitrary ID$M^N_O$ for criticality is a subtle business that we handle using an exhaustive computer algorithm.  Finding a more compact analytic method remains an open problem.

Determining the total number of representative IDs for a given $M$ and $N$ is a computationally difficult problem.  Our exhaustive search method is guaranteed to find all possible types of IDs, but it also finds many permutations of the same representative ID.  In general, to compare two canonical ID$M^N$s and determine if they are truly distinct, one must take all $N! M!$ permutations of the order of the qubits and observables in one ID and compare these to the other ID, though in some cases there are significant shortcuts for this computation.  To give a sense for the scope of this problem, the exhaustive search returns 163,910 different critical ID$6^6$s, every pair of which must be checked to determine if they belong to the same representative ID, and each of those checks requires as many as $(6!)^2$ permutations.  Perhaps thankfully, we cannot even complete our exhaustive searches for any IDs larger than $M=6$ and $N=7$, and indeed we find a staggering 2,466,888 different critical ID$6^7$s to sort through.  So the question of how many representative critical IDs exist for larger values of $M$ and $N$ remains open.

\begin{table}[h!]
\caption{Each entry shows the number of representative ID$M^N_O$s for given values of $M$, $N$, and $O$, with $O$ given by the subscript.  For example, there is one representative ID$5^4_0$ and there are 6 representative ID$5^4_2$s.  All IDs have been exhaustively enumerated for sizes to the left of the dividing line (though the exact number of representative IDs is not yet known for all these cases).  To the right of the line are sizes for which examples are known, but no complete search has been performed.}
\begin{center}
\begin{tabular}{cc|cccccc}
    \multicolumn{2}{c}{} & \multicolumn{6}{c}{$M$}\\
    & & 3 & 4&5 & 6&7 & 8 \\\cline{2-8}
    \multirow{16}{*}{$N$}&2 & $1_2$ &  &  & \multicolumn{1}{c|}{} &  &  \\
    &3  &  & $1_0$,$1_2$ &  & \multicolumn{1}{c|}{} &  &  \\
    &4  &  & $1_2$,$1_4$ & $1_0$,$6_2$ & \multicolumn{1}{c|}{} &  &  \\
    &5  &  &   & $1_0$,$22_2$,$5_4$ & \multicolumn{1}{c|}{$22_0$,$170_2$,$18_4$} &  &  \\
    &6  &  &  & $2_0$,$51_2$,$20_4$,$1_6$ & \multicolumn{1}{c|}{$x_{0\ldots 6}$} & $x_{0\ldots 6}$ &  \\
    &7  &  &  & $62_2$,$45_4$,$2_6$  & \multicolumn{1}{c|}{$x_{0\ldots6}$} & $x_{0\ldots 6}$ & $x_{0\ldots 4}$ \\\cline{6-6}
    &8  &  &  & \multicolumn{1}{c|}{} & $x_{0\ldots8}$ & $x_{0\ldots 8}$ &  $x_{0\ldots6}$ \\
    &9  &  &  &\multicolumn{1}{c|}{}  & $x_{0\ldots8}$ & $x_{0\ldots8}$ & ? \\
    &10  &  &  &\multicolumn{1}{c|}{}  & $x_{0\ldots10}$ & $x_{0\ldots10}$ & ? \\
    &11  &  &  &\multicolumn{1}{c|}{}  & $x_{2\ldots10}$ &$x_{0\ldots10}$  & ? \\
    &12  &  &  &\multicolumn{1}{c|}{}  & ? & $x_{4\ldots12}$ & $x_{12}$ \\
    &13  &  &  &\multicolumn{1}{c|}{}  &  & $x_{6\ldots12}$ & ? \\
    &14  &  &  &\multicolumn{1}{c|}{}  &  & ? & ? \\
    &15  &  &  &\multicolumn{1}{c|}{}  &  & $x_{14}$ & ? \\
    &16  &  &  &\multicolumn{1}{c|}{}  &  & $x_{16}$ & ? \\
     &17  &  &  &\multicolumn{1}{c|}{}  &  & ? & ? \\
\end{tabular}
\end{center}
\label{TableMN}
\end{table}

For brevity we refer the reader to our other works to see example families of critical IDs for arbitrarily large values of $N$ \cite{WaegellThesis, WA_Nqubits, WA_4qubits}.

Every type of representative critical ID belongs to at least one graph stabilizer group, which in turn shows that there exist permutations of every ID that belong to a fully real-valued subgroup of the $N$-qubit Pauli group - specifically the subgroup in which $Y_q$ appears for an even number of qubits in every observable in the group.  This means that no representative IDs are lost if we restrict ourselves to rebit states (real-valued states) \cite{caves2001entanglement, wootters2001entanglement}.  However some representative critical KS sets are lost by this restriction (see Figure 2 of \cite{WA_3qubits} for an example).

Another interesting feature of IDs is that for $N\geq4$ qubits there exist ID$M^N$ with $M < N+1$, whose joint eigenbasis consists of maximally entangled projectors with rank, $r = 2^{N-M+1}$, greater than one.  The projectors in such an eigenbasis can also be seen as partitioning the space $\mathbb{C}^{2^N}$ into $2^{M-1}$ subspaces $\mathbb{C}^r$.  Every state within one of these subspaces is maximally entangled, a remarkable feature that could have interesting applications.  Table \ref{TableMN} shows an overview of known sizes of critical IDs, as well as counts for those we have fully enumerated and sorted.\footnote{Because they cannot be used in critical NKS or KS sets, we excluded all Positive ID$M^N_0$ from our original searches, and thus also from the listings of critical IDs throughout this paper.  Their anticommutation structures may nevertheless be useful for other applications, and so their existence should not go unmentioned.}

A very interesting open question is to identify the lower bound $L(N)$ on $M$, for the existence of critical ID$M^N$s (i.e. such that $L(N) \leq M \leq N+1$).  We have determined some of the lower values of $L(N)$ from exhaustive numerical searches, and from some other special examples.  These are $L(2)=3$, $L(3)=4$, $L(4) = 4$, $L(5,6,7)=5$, $L(8,9,10,11,\ldots$?$) = 6$, $L(12,13,14,15,16,\ldots $?$) = 7$.  Clearly the dimension, $d=2^{N- L(N) +1}$, of the largest available maximally entangled subspace grows quite rapidly with $N$.  This growth also enables particularly compact tests of the KS theorem for systems of many qubits through a family of KS sets we call Kites, which we have described in detail elsewhere \cite{WaegellThesis, WA_Nqubits}.

Table \ref{SmallIDs} shows our maximal example IDs for $L(N)=5,6,7$.

For a quite different discussion of the geometry of stabilizer codes of the $N$-qubit Pauli group, see \cite{Bierbrauer}.

\subsubsection{SQP Graphs and ID Color-Hypergraphs}\label{IDGraphs}

\begin{figure}[h!]
\centering
\qquad
\subfloat[][]{
\includegraphics[width=.75in]{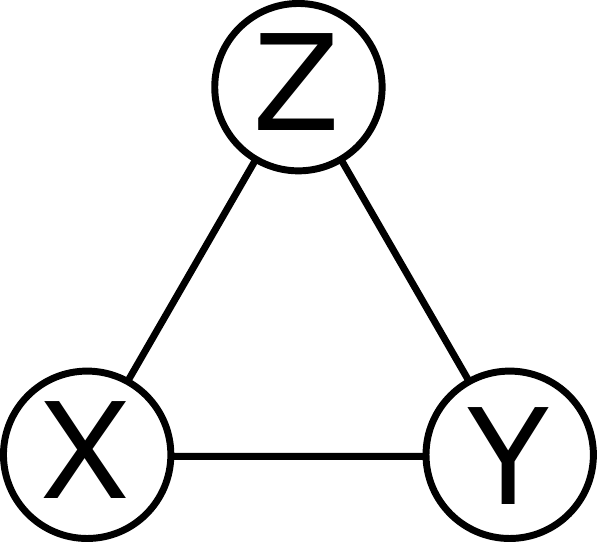}
\label{SQPG3}}
\qquad
\subfloat[][]{
\includegraphics[width=.75in]{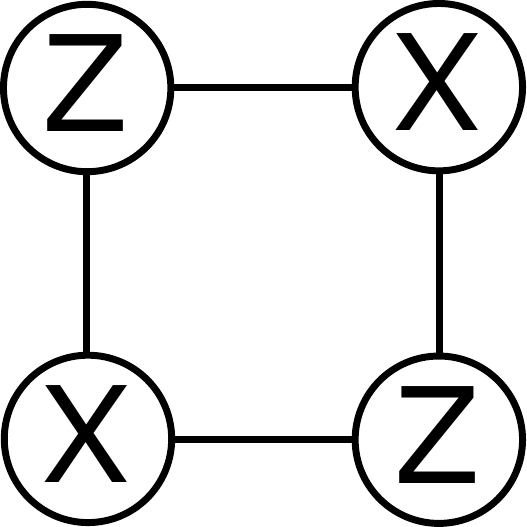}
\label{SQPG4}
}
\qquad
\subfloat[][]{
\includegraphics[width=.75in]{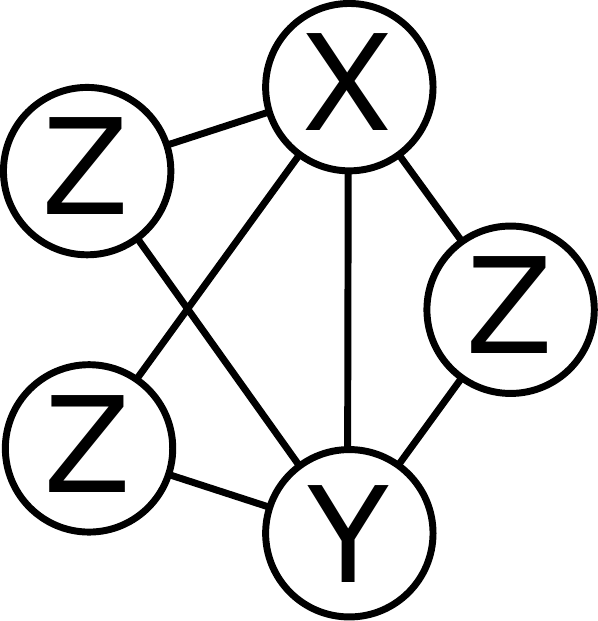}
\label{SQPG5}
}
\qquad
\subfloat[][]{
\includegraphics[width=.75in]{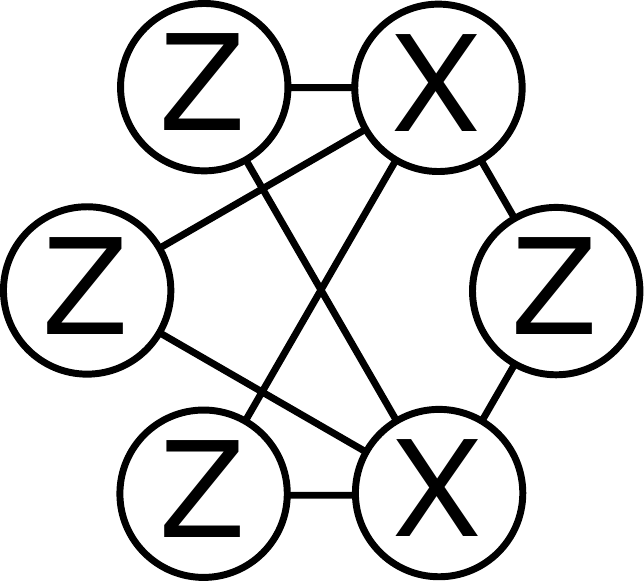}
\label{SQPG6_1}
}
\qquad
\subfloat[][]{
\includegraphics[width=.75in]{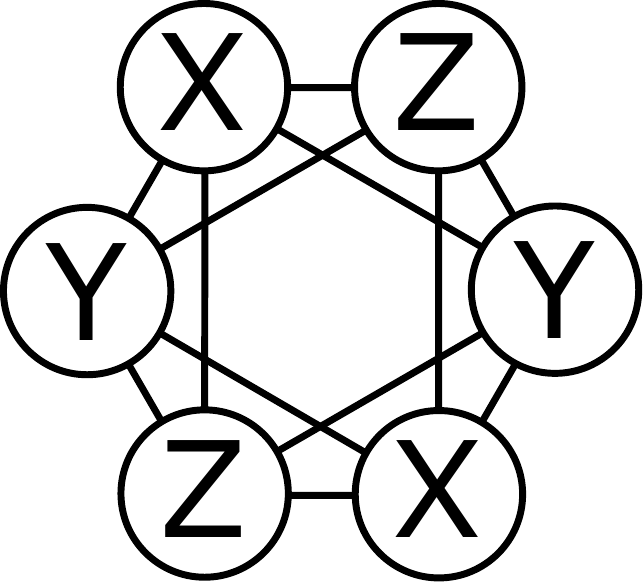}
\label{SQPG6_2}
}
\qquad
\subfloat[][]{
\includegraphics[width=.75in]{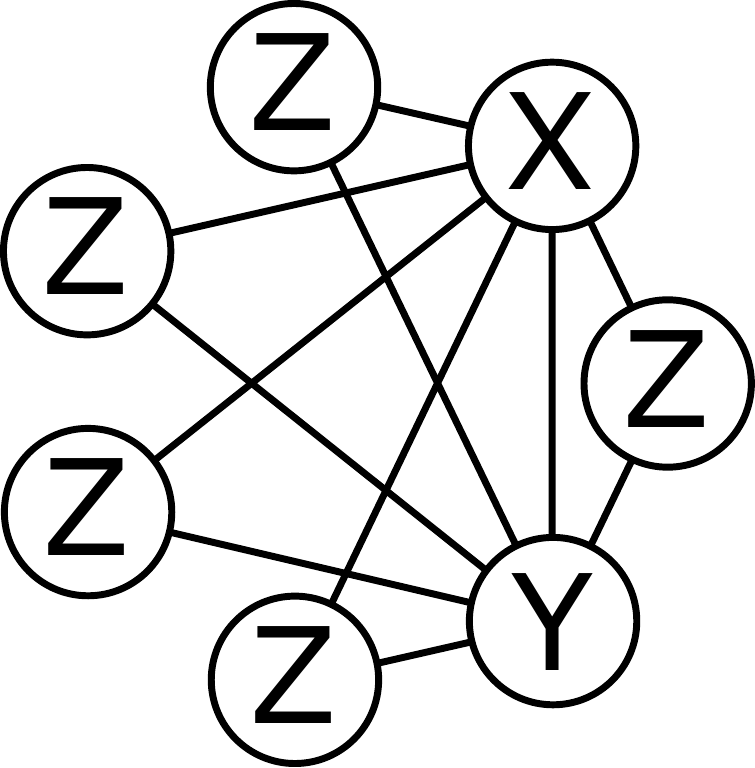}
\label{SQPG7_1}
}
\qquad
\subfloat[][]{
\includegraphics[width=.75in]{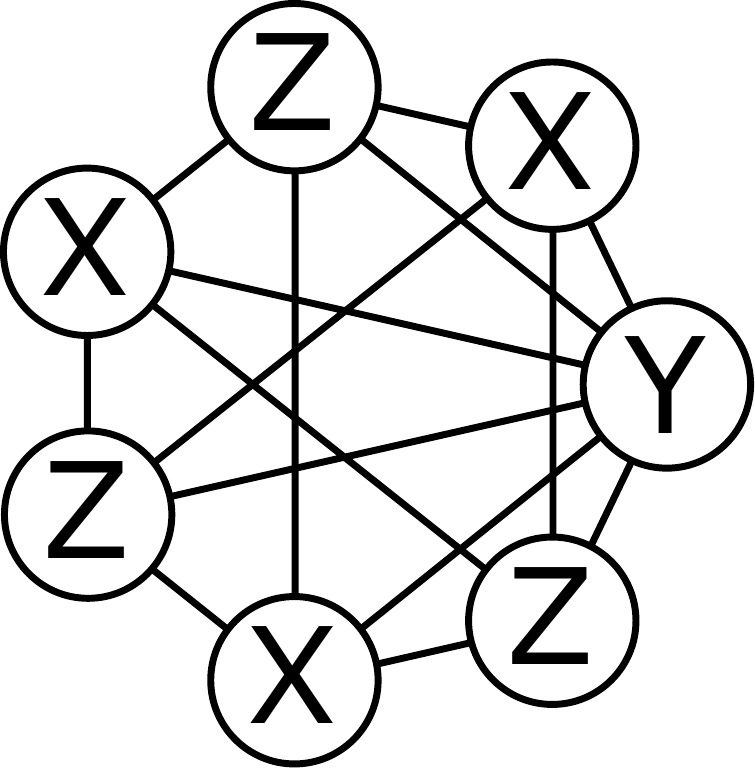}
\label{SQPG7_2}
}
\qquad
\caption{All nonisomorphic anticommutation graphs for Nontrivial SQPs with up to 7 Pauli observables.}\label{SQPgraphs}
\end{figure}

We can restate most of the discussion of SQPs and IDs in a graph-theoretic framework that will lend us some useful concepts and terminology.  To begin we define an anticommutation graph for each SQP.  This is a graph whose vertices represent Pauli observables $\{Z,X,Y\}$ and whose edges join anticommuting observables.  Each SQP graph always has an even number of edges at each vertex, which follows from the definition of an SQP.  Trivial SQP graphs have no edges at all.  The number of vertices in one of these graphs is simply the number of Pauli observables in the SQP, and an arbitrary number of disconnected vertices can be added for elements $I$ in the SQP.  It is important to note that we will not make any attempt to order the vertices of an SQP graph, and so information about the signs of SQPs and IDs in this graphical representation will be somewhat obscured, while the underlying entanglement structure will be made more apparent.  The simplest SQP graphs are shown in Figure \ref{SQPgraphs}.

To form an ID for multiple qubits, we will begin by taking several SQP graphs (each for a different qubit) and assigning each one a different color.  These graphs are built on a common set of vertices, and so we obtain a color-hypergraph, that in general can have multiple edges of different colors connecting the same pair of vertices, which is to say edges with multiple colors.  Now we can restate some definitions in terms of these graphs.

The color-hypergraph is an ID iff each color is a legal SQP graph and every edge has an even number of colors.

Two IDs belong to the same representative ID iff their color-hypergraphs are isomorphic.

An ID is critical iff it is impossible to delete any subset of vertices (along with associated edges) and/or colors (i.e. qubits) from the color-hypergraph such that the remaining color-hypergraph still satisfies the above requirements to be an ID.

\begin{figure}[h!]
\centering
\qquad
\subfloat[][]{
\includegraphics[width=1.25in]{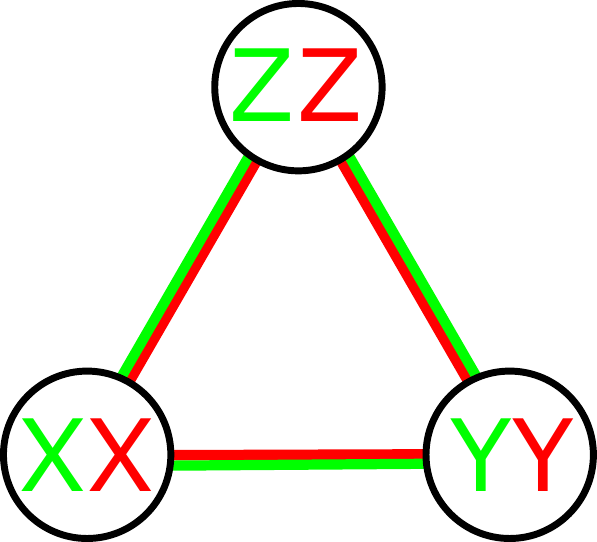}
\label{IDG2}}
\qquad
\subfloat[][]{
\includegraphics[width=1.25in]{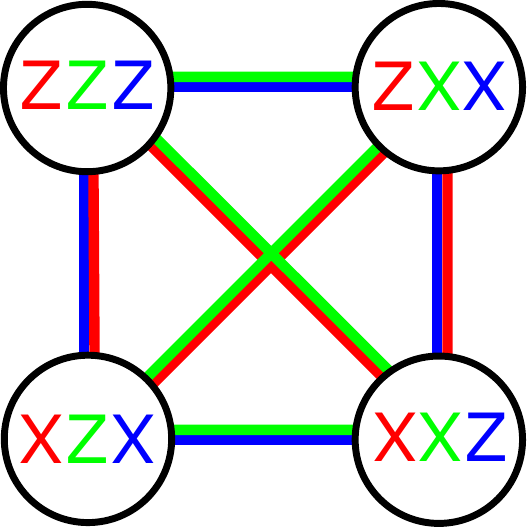}
\label{IDG3_1}
}
\qquad
\subfloat[][]{
\includegraphics[width=1.25in]{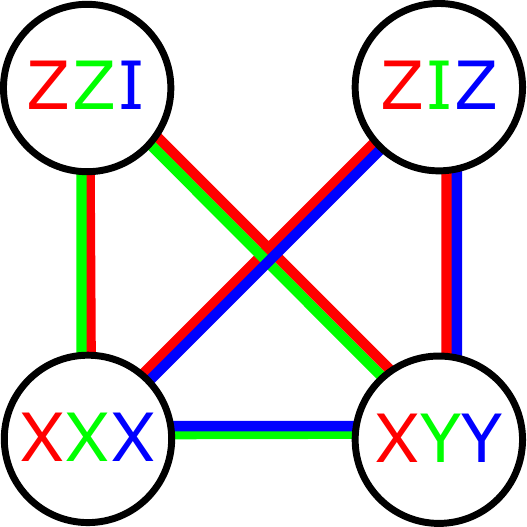}
\label{IDG3_2}
}
\qquad
\caption{(Color online) Color-hypergraphs for all representative critical IDs with up to 3 qubits.  The observables need not be shown at the vertices of the hypergraph, but this is done here to clarify the example.  The important point is that all possible permutations of the ID will yield isomorphic color-hypergraphs, and thus distinct IDs must have nonisomorphic color-hypergraphs.}\label{IDGraphs1}
\end{figure}

The color-hypergraphs for a few of the simplest IDs are shown in Figures \ref{IDGraphs1} and \ref{IDGraphs2}.

It is our hope that the machinery of graph theory can be applied to some of the problems and open questions that were discussed above.

\subsection{NKS Sets}\label{NKS}

Critical NKS sets are sets of one or more IDs as described above.  The IDs in an NKS set need not be critical in the sense given for IDs, but instead the NKS will be called critical if no subset of IDs and/or qubits can be removed from the set such that a smaller NKS set remains.  These critical NKS sets are then the minimal structures in the $N$-qubit Pauli group that demonstrate $N$-qubit contextuality.

An odd number of the IDs in an NKS set must be negative, while each single-qubit Pauli observable must appear an even number of times among those IDs.  Negative IDs composed of only Even SQPs are thus already single-ID NKS sets (see Tables \ref{q3ID1} and \ref{ID5_q4_1}), and are thus critical iff the ID is critical.  Positive IDs composed of only Even SQPs play no role in critical NKS sets, since no permutation of the SQPs will change the sign of the ID, and it can always be discarded.  Odd SQPs on the other hand, can always change sign under reflection of their coordinate system, and so IDs containing Odd SQPs can always be used in multi-ID NKS sets.  Any set of IDs that each contain Odd SQPs, with an odd number of them permuted to be negative, and such that an even number of the IDs have an Odd SQP for each qubit, is automatically an NKS set - though it is not necessarily critical.  The simplest NKS set of this type is given in Table \ref{NKS2}.
\begin{table}[h!]
\begin{center}
\qquad
\subfloat[][]{
\begin{tabular}{cc}
$Z$ & $Z$\\
$X$ & $X$\\
$Y$ & $Y$\\
\end{tabular}\label{NKS2_1}}
\qquad
\subfloat[][]{
\begin{tabular}{cc}
$Z$ & $X$\\
$X$ & $Z$\\
$Y$ & $Y$\\
\end{tabular}\label{NKS2_2}}
\qquad
\end{center}
\caption[]{These two critical ID$3^2_2$ form the simplest critical mutli-ID NKS set, using the 2-qubit CNS of Table \ref{CKS2}.  The ID of \subref{NKS2_1} is negative, and a simple permutation of the second SQP gives the positive ID of \subref{NKS2_2}.  Between the two IDs, each single-qubit Pauli observable $\{Z_q,X_q,Y_q\}$ appears twice.}
\label{NKS2}
\end{table}

If a set of IDs are each critical over their Odd SQPs (meaning that those specific qubits cannot be removed to leave a smaller nontrivial ID), and also in the sense that it cannot be separated into the product of two or more smaller IDs, then they can be used to form critical NKS sets using structures that we call Critical NKS Structures \footnote{In  \cite{WaegellThesis}, CNSs were called `Composite Kernel Structures (CKSs),' but since we have dropped the name `Kernel,' we have now adopted CNS for these structures.} (CNSs for short).  CNSs are built from generalized IDs.  In a generalized ID, we reduce each Odd SQP to a single `O' and each other SQP to a single `I'.  For example, the ID$4^3_2$ of Table \ref{q3ID2} can be reduced to the generalized ID `IOO,' since it is critical over its Odd SQPs, and it is not the product of two smaller IDs.  Generalized IDs always have an even number of Odd SQPs, since this is a general property of IDs.  Each generalized ID is a qubit-ordered set, in the sense that it is the relative order of the generalized IDs in a CNS that is important.  The overall order of the qubits in a complete CNS is of course arbitrary.

We depict CNSs as tables in which each row is a different generalized ID, and each column contains an even number of `O's (the requirement for the set of IDs to form an NKS set).  Furthermore, a CNS is by definition critical in the sense that it is impossible to discard any generalized IDs and/or qubits such that a smaller CNS remains.  The representative CNSs are a special class of geometric objects for $O$ Odd qubits.  All 16 representative CNSs for up to $O=5$ are shown in Table \ref{CNSs}.  We have also enumerated all 109 representative CNSs for $O=6$, and all 1,521 for $O=7$, and these are presented on the website as well \cite{MainWebsite}.

All ID$M^N_O$s with a given value of $O$ reduce to the same generalized ID (up to an arbitrary number of `I's and permutations of qubit order), which means that any of these IDs can be used interchangeably in a given CNS to give an NKS set - provided that no two IDs are identical, and an odd number of them are negative.  Indeed {\it every} such combination of IDs is a critical NKS set, meaning that the number of distinct NKSs that can be formed in this way grows extraordinarily quickly with $N$.  The `I's in most CNSs represent either Trivial or Even SQPs in the pre-generalized IDs, and so it is also easy to see that the individual IDs need not be fully critical in order for the CNS to lead to a critical NKS set - they must only be critical over their Odd SQPs.  In some cases, not all of the IDs even need to be critical over their Odd SQPs in order for the NKS to be critical.  Checking the criticality of an arbitrary NKS set is a subtle business that we handle with an exhaustive computer algorithm, and finding  a more compact analytic method remains an open problem.

One particularly transparent family of CNSs are what we call the $N$-qubit Rings, with only a single pair of Odd SQPs assigned to each qubit and each generalized ID, in such a way that all $N$ qubits are linked in a single closed loop.  Tables \ref{CKS2}, \ref{CKS3}, \ref{CKS4_4_2}, and \ref{CKS5_5_6} belong to this family, and can be used to give what we have elsewhere called the Wheel and Whorl family of KS sets \cite{WaegellThesis, WA_Nqubits}.

Another simple family for even values of $O$ is simply a pair of generalized IDs with all Odd SQPs.  Tables \ref{CKS2} and \ref{CKS4_2} belong to this family, which can be used to give what we have elsewhere called the Kite family of KS sets \cite{WaegellThesis, WA_Nqubits, WA_3qubits}.

The last important point is that for IDs that are critical over Even SQPs, the Even SQPs need not be assigned to the $O$ qubits of the CNS, and can instead be assigned to additional qubits, which are then inexorably linked to that NKS set.  In this way, CNSs for $O$ odd qubits can be used to obtain critical NKS sets for far more than $O$ qubits.

For example, a simple permutation of the critical ID$6_2^{11}$ in Table \ref{M6N11}, paired with the original, gives the two generalized IDs for a 20-qubit critical NKS set.  We again use the 2-Odd-qubit CNS of Table \ref{CKS2}, and we assign the 9 Even SQPs from each ID to 18 different qubits, with Trivial SQPs for those qubits in the partner ID (so we truly have two noncritical ID$6_2^{20}$s of opposite sign), as shown in Table \ref{NKS20}.  In fact, we can use permutations of the same ID along with the Ring CNS for any $O\geq2$ to obtain an NKS set for $N=10O$ qubits in just the same way.
\begin{table}[h!]
\centering
\subfloat[][]{
\begin{tabular}{cccccccccccccccccccc}
O & O & E & E & E & E & E & E & E & E & E & I & I & I & I & I & I & I & I & I \\
O & O & I & I & I & I & I & I & I & I & I & E & E & E & E & E & E & E & E & E \\
\end{tabular}}
\caption[]{This shows how the Odd and Even SQPs of 2 critical ID$6_2^{11}$s are permuted to form a 20-qubit critical NKS set.}\label{NKS20}
\end{table}

\subsubsection{CNS Color-Hypergraphs}

Unsurprisingly each of the CNSs can also be depicted as a color-hypergraph, with nonisomorphic hypergraphs for each representative CNS.  Again, we will begin by constructing a graph for each qubit.  The vertices of the graph will correspond to the elements `O' and `I' of the generalized IDs, and the edges will connect every pair of `O's (since there is no favored sub-pairing for a set of 4 or more `O's).  We may also have an arbitrary number of disconnected vertices for elements `I'.  We will call these Generalized Qubit graphs, and they are simply the family of complete graphs on an even number of vertices, plus any additional number of disconnected vertices.

To form a CNS for multiple qubits, we will begin by taking several Generalized Qubit graphs (each for a different qubit) and assigning each one a different color.  These graphs are built on a common set of vertices, and so we obtain a color-hypergraph, that in general can have multiple edges of different colors connecting the same pair of vertices, which is to say edges with multiple colors - in a manner identical to what we did for the ID color-hypergraphs.  Now we can restate some definitions for CNSs in terms of these hypergraphs.

The color-hypergraph is a CNS iff every color is a legal Generalized Qubit graph and an even number of colors meet at each vertex.

Two CNSs belong to the same representative CNS iff their color-hypergraphs are isomorphic.

A CNS is critical iff it is impossible to perform deletions of vertices and/or colors on its color-hypergraph according to the following rules in order to obtain a smaller (legal) CNS color-hypergraph.  The deletion rules are 1) If we remove a vertex (generalized ID), we must also remove all edges associated to that vertex. 2) If we remove the last edge of a given color, either by deleting an entire color (qubit), or as a result of rule 1, then we must also remove all vertices that color was connected to (this is because generalized IDs are entangled over all of their odd qubits, i.e. `O's, by definition.). It should be clear that an application of rule 1 may force an application of rule 2, and vice versa, in a chain of deletions.

\begin{figure}[h!]
\centering
\qquad
\subfloat[][]{
\includegraphics[width=1.25in]{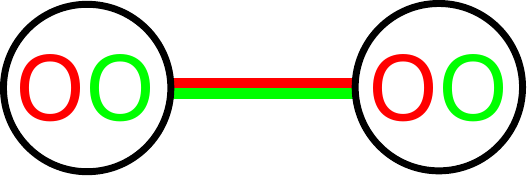}
\label{CNSG2}
}
\qquad
\subfloat[][]{
\includegraphics[width=1.25in]{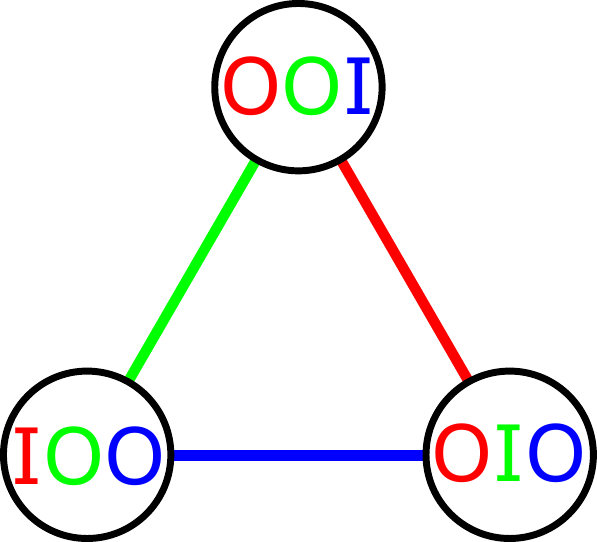}
\label{CNSG3}
}
\qquad
\subfloat[][]{
\includegraphics[width=1.25in]{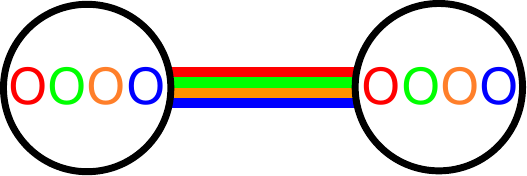}
\label{CNSG4_1}
}
\qquad
\subfloat[][]{
\includegraphics[width=1.25in]{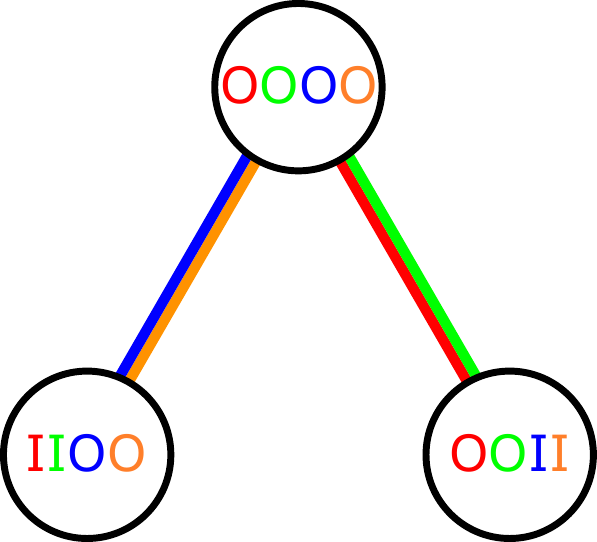}
\label{CNSG4_2}
}
\qquad
\subfloat[][]{
\includegraphics[width=1.25in]{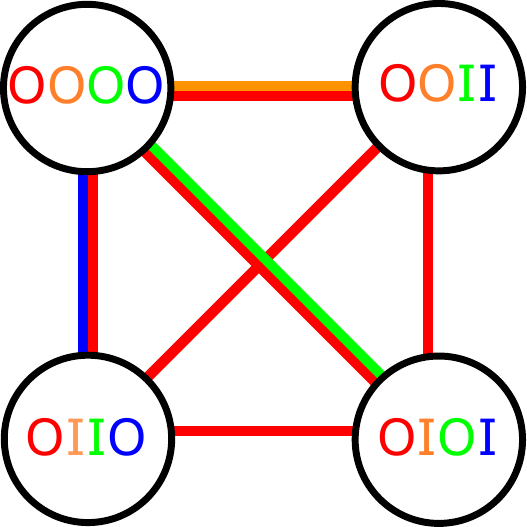}
\label{CNSG4_3}
}
\qquad
\subfloat[][]{
\includegraphics[width=1.25in]{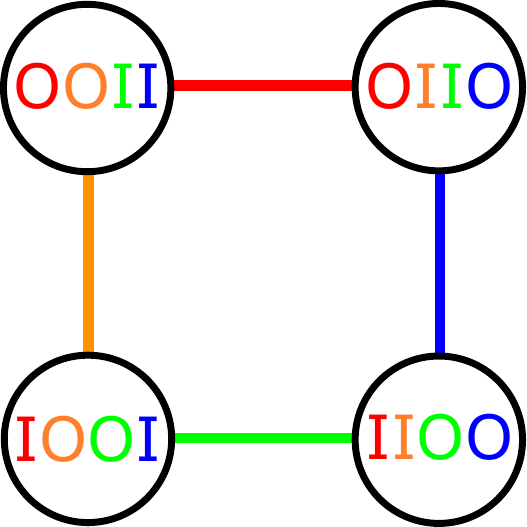}
\label{CNSG4_4}
}
\qquad
\caption{(Color online) Color-hypergraphs for all representative CNSs with up to 4 Odd qubits (see Table \ref{CNSs}).  The generalized IDs need not be shown at the vertices of the hypergraph, but this is done here to clarify the example.}\label{CNSgraphs}
\end{figure}

The CNS color-hypergraphs $O = 2,3,4$ are shown in Figure \ref{CNSgraphs}.

While the task of enumerating all of the representative CNSs for a given value of $O$ is not as computationally difficult as it is with IDs, it still presents a challenge for larger values of $O$. Perhaps the machinery of graph theory can provide us with a more efficient method.

\subsection{KS Sets}

Critical KS sets are sets of IDs as described above.  Again, the IDs in a KS set need not be critical in the sense given for IDs, but instead the KS will be called critical if no subset of IDs and/or qubits can be removed from the set such that a smaller KS set remains.  These critical KS sets are then the minimal structures in the $N$-qubit Pauli group that demonstrate quantum contextuality.

Every critical NKS set can be used (in a number of ways) to generate a critical KS set.  For example, both the Peres-Mermin Square and the 2-qubit Whorl (Figures 1 and 2 of \cite{WA_3qubits}) are both built up from the same NKS set.  In analogy to the relationship between ID-criticality and NKS-set-criticality, it is also possible to use noncritical NKS sets to build critical KS sets (see Figure 9 of \cite{WaegellThesis} for a specific example).  It appears to be true that every KS set of IDs contains at least one NKS set of IDs as a subset, and we conjecture that this is always the case.

We present here one such generation method which always uses a critical NKS set to obtain a critical KS set \cite{WaegellThesis}.  To get from the NKS set to the KS set, we need to add some new IDs.  The new IDs will be constructed entirely out of Trivial SQPs and so they will all be positive.  This guarantees that an odd number of the IDs in the KS set will be negative, since this was already true for the NKS set.  For each observable that appears in an odd number of IDs from the NKS set, we will generate one new ID, by multiplying that observable by its own single-qubit decomposition.  For example, an observable $ZZXZ$ would be combined with the new observables $ZIII$, $IZII$, $IIXI$, $IIIZ$ to form a new positive ID$5^4$.  This guarantees that every observable will appear in an even number of IDs in the KS set, since the unpaired observables from the NKS set have been paired by the new IDs, and each new single-qubit observable already appeared in an even number of the IDs in the NKS set.

For clear examples, note that the Peres-Mermin Square \cite{Mermin_SquareStar, WA_24Rays}, which is a KS set, contains the NKS set of Table \ref{NKS2}, while the GHZ-Mermin Star \cite{Mermin_SquareStar, WA_3qubits} contains a permutation of the single-ID NKS set of Table \ref{q3ID1}.  Both of these KS sets are obtained from their NKS subset by applying the method described above.

These and many other interesting examples of KS sets that can be obtained in this way are given in convenient hypergraph diagrams in several of our other works \cite{WaegellThesis, WA_Nqubits, WA_4qubits, WA_3qubits}, as well as some cases given by alternate generation methods.

Every KS set also gives rise to a saturated set of projectors and bases which can also be used to prove the Kochen-Specker theorem (these sets of projectors are themselves called `KS sets' in most of the literature on this topic).  These saturated sets always contain critical parity proofs of the Kochen-Specker theorem as subsets, an issue we have explored in great detail elsewhere \cite{WaegellThesis, WA_Nqubits, WA_3qubits, WA_24Rays, WA_60Rays}.

The last important piece of the nonclassicality puzzle is nonlocality, but unsurprisingly it turns out that KS sets are already recipes for experimental tests of both quantum contextuality and quantum nonlocality \cite{aravind2002bell}.  Single-ID NKS sets are also recipes for tests of quantum nonlocality in a somewhat independent manner \cite{GHSZ, Walther2005}.  Both of these generalizations have been explored elsewhere in detail, and so we do not recount them here \cite{WaegellThesis}.

We should however discuss the subtle distinction between the criticality of Single-ID NKS sets and what has been called the {\it (weak) genuineness of $n$-partite $d$-level} GHZ paradoxes \cite{YuOhGraphStates}, for the case of $N\equiv n$ and $d=2$ (i.e. the $N$-qubit Pauli group).  It is possible for a genuine GHZ paradox to be demonstrated using a Single-ID NKS set, even if the ID is not critical, provided that the prepared eigenstate of the ID is a genuine $N$-qubit maximally entangled state (this is the case in \cite{Walther2005}, and in some of the examples given in \cite{YuOhGraphStates} for $d=2$).  If the ID is not critical, then there also exist alternate product eigenstates that need only contain a maximally entangled subset of $\eta$ qubits ($3 \leq \eta < N$), and which still demonstrate the GHZ paradox (and some of the parties can simply be ignored).  This happens because a noncritical ID automatically belongs to stabilizer groups of several different degrees of entanglement.  We should stress here that we are not saying that the GHZ paradoxes given by these authors are flawed; they do indeed show genuine $N$-partite nonlocality.  What we are saying is that the identical set of measurements would still show $\eta$-partite nonlocality if a different eigenstate were prepared for the experiment.  If however, the ID is critical, then the previous statement no longer holds, and only genuine $N$-partite nonlocality can be demonstrated by that set of measurements, and only using a genuine $N$-qubit maximally entangled eigenstate.  We will call these cases {\it strongly genuine $N$-partite $2$-level} GHZ paradoxes to distinguish them.  We presented a family of strong GHZ paradoxes for all even $N \geq 4$ in \cite{WA_4qubits} \footnote{We take this opportunity to correct a mistaken comment we made in this paper.  We said that the states we were discussing were not Graph States because of the structure of the specific IDs, but now that we are more familiar with graph stabilizer groups, it is obvious that the eigenstates of all critical IDs are Graph States.}, and the family presented for all odd $N\geq5$ in \cite{aravind2002bell} is also strong.  Table \ref{TableMN} shows the overview of other strong GHZ proofs we have identified - these are just the IDs with $O=0$.  It is also noteworthy that the 4-qubit GHZ state cannot be used to give a strong GHZ paradox, while the 4-qubit Cluster state can, which can be seen by looking at Tables \ref{q4ID_1} and \ref{ID5_q4_1}.

As we have already discussed, noncritical Single-ID NKS sets are still of very limited use in constructing critical ({\it genuine $N$-qubit}) KS sets.  Specifically, only strong GHZ paradoxes can be used to trivially generate critical KS sets using the method described above (though they could still be used in a nontrivial method).  Because the KS theorem is state-independent, the structure of a KS set depends entirely on the entanglement (criticality) structure of the IDs in the set, rather than on the entanglement of any specific state.

\subsection{Conclusions}

We anticipate that all of these minimal objects we have classified will find numerous applications, since they seem to capture exactly the nonclassical physics that we wish to exploit as a resource for quantum information processing with $N$ qubits.  They can be used directly for tests of entanglement, (N-qubit) contextuality, and nonlocality, and they also have applications for quantum key distribution \cite{ekert1991quantum}, parity oblivious transfer \cite{bennett1992practical}, quantum error correction \cite{bennett1996mixed, calderbank1997quantum, steane1996multiple}, quantum dimension certification \cite{CabelloKSDim}, quantum algorithms \cite{briegel2001persistent, raussendorf2001one, anders2009computational, cleve1998quantum, ekert1998quantum}, etc...
\begin{figure}[h!]
\begin{center}
\includegraphics[width=4in]{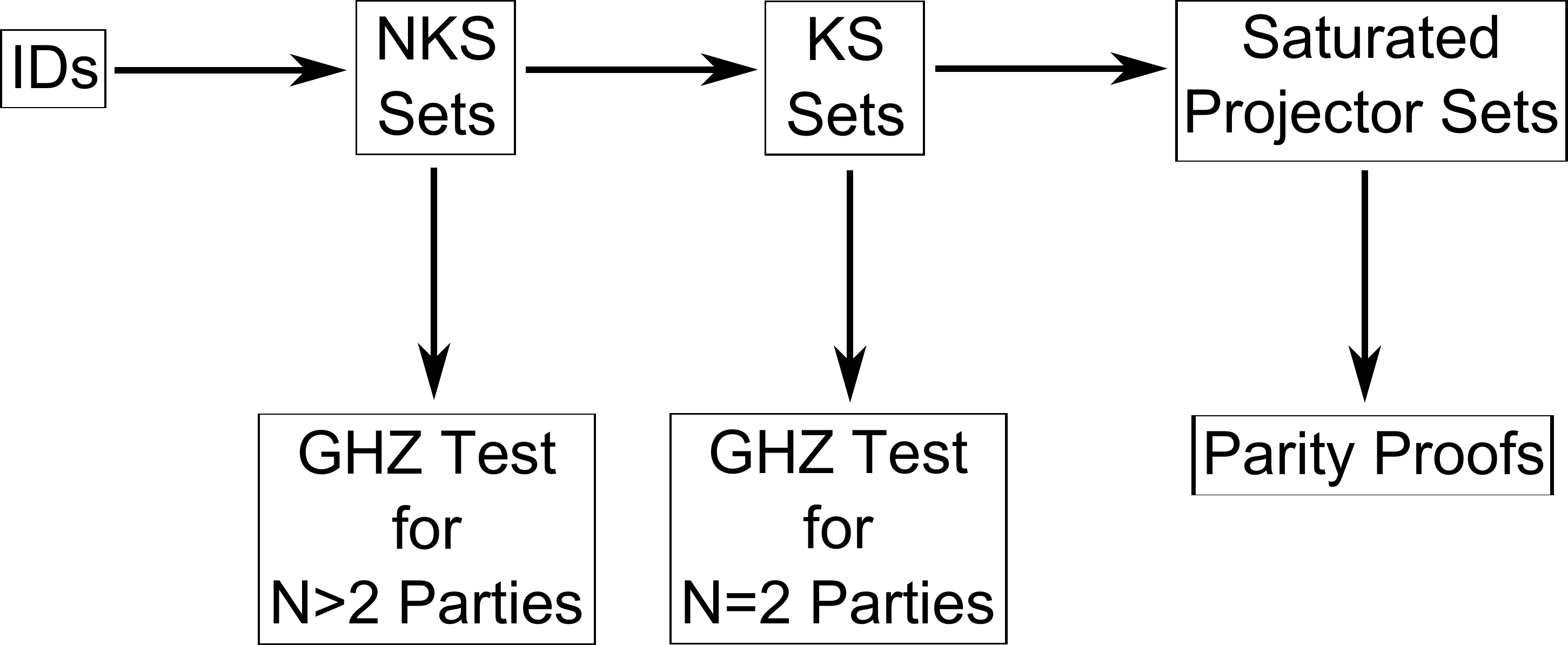}
\caption{The basic progression of nonclassical phenomena in the $N$-qubit Pauli group, showing the order in which each type of nonclassical structure is built up from the previous ones.}\label{FlowChart}
\end{center}
\end{figure}

We also hope by now to have conveyed the overall picture of how the various manifestations of nonclassicality within the $N$-qubit Pauli group are interconnected, and how they follow a sort of progression that begins with IDs, which are the fundamental objects of entanglement, and then moves to NKS sets, and then to KS sets, which are the fundamental objects of contextuality and ultimately nonlocality.  Figure \ref{FlowChart} illustrates this progression.  It should also be noted that with systems of qubits, all of this nonclassicality follows directly from entanglement between the qubits, and none of the qubits in a critical structure ever fail to be entangled with at least one other qubit in that structure.

As a final aside, consider that if we were to derive a theory of quantum mechanics such that every discrete Hilbert space is composed of some collection of qubits - possibly with some internal states `frozen out' - then it would follow that all nonclassicality in these discrete spaces arises because of entanglement.  Specifically this addresses the issue that a spin-1 particle can be used to show quantum contextuality \cite{Klyachko, peres1993quantum}, seemingly without entanglement.  If we suppose that the spin-1 particle is fundamentally composed of 2 qubits, with the singlet state frozen out, then we can see that those 2 qubits must be entangled for the proof of contextuality to work.  Such a qubit-based derivation of quantum mechanics would then reduce the entire discussion of nonclassicality to a discussion about entanglement, while also describing everything in terms of the fundamental unit of quantum information.\newline

\textbf{Acknowledgements:}  I would like to thank P.K. Aravind for many useful discussions as I pursued this research, and he was directly involved in the various precursor projects to which I have referred throughout.  I also thank Walter Lawrence for useful discussions.

\bibliographystyle{ieeetr}
\bibliography{id_paper.bbl}
\pagebreak

\subsection{Appendix}
\subsubsection{}

We have made the assertion that if the observables in an ID cannot be factored into the direct product of two sets of mutually commuting observables, then it follows that the joint eigenbasis of this ID must be composed of maximally entangled states.  This is actually just a special case of a more general result that relates any set of commuting multipartite observables to the entanglement between their parts.  Here we give a formal derivation for this alternate definition of entanglement.

Let $\{A_m\}$, $m=1,\ldots,M$, be a set of $M$ mutually commuting multipartite observables, and let $\{|\psi_i\rangle\}$, $i = 1,\ldots,\frac{d}{r}$, be a spanning joint eigenbasis of $\{A_m\}$ composed of rank $r$ projectors, such that
\begin{equation}
A_m |\psi_i\rangle = \lambda_{mi} |\psi_i\rangle.
\end{equation}
Now, suppose that each element of $\{|\psi_i\rangle\}$ can be factored into the direct product of states from two different subparts of the system,
\begin{equation}
|\psi_i\rangle = |\phi_i\rangle \otimes |\chi_i\rangle.
\end{equation}
It follows that $\{A_m\}$ must also be factorable into the same parts, so that
\begin{equation}
A_m = B_m \otimes C_m,
\end{equation}
and the $\{|\phi_i\rangle\}$ ($\{|\chi_i\rangle$\}), which may now contain repeated elements, must be a spanning set of joint eigenstates (i.e. an eigenbasis) of the set of observables $\{B_m\}$ ($\{C_m$\}), which may also now contain repeated elements, corresponding to projectors of rank $s$ ($t$), such that $r=st$, and
\begin{equation}
B_m |\phi_i\rangle \otimes C_m |\chi_i\rangle = \sigma_{im} |\phi_i\rangle \otimes \rho_{im}|\chi_i\rangle.
\end{equation}

Therefore, if $\{A_m\}$ cannot be factored (in any way) into the direct product of two subparts, such that $\{B_m\}$ and $\{C_m\}$ are each mutually commuting sets, then it follows that the $\{|\psi_i\rangle\}$ are maximally entangled states of the multipartite system.  We then say that $\{A_m\}$ is an entangled set of observables.  Furthermore, if removing any one observable from $\{A_m\}$ leaves behind a set that is no longer entangled, we say that $\{A_m\}$ is minimal.

Now, if $\{A_m\}$ is an entangled $N$-qubit ID with the additional property that its observables cannot be separated into two smaller $N$-qubit IDs, then we say that it is critical.

Within the $N$-qubit Pauli group, an ID is obtained from any entangled set of free generators, by adding a final observable to the set equal to the overall product of those generators, and all minimal entangled sets of free generators within the $N$-qubit Pauli group give rise to critical IDs.  To see this, consider that for any critical ID$M^N$, one obtains $M$ different minimal free sets of entangled generators by removing any one of the $M$ observables from the ID, and that the critical ID is of course restored by returning the observable, which is by definition equal to the overall product of the other $M-1$ observables.  If the set of free entangled generators is not minimal, then it generates a noncritical ID, whose observables can be separated into multiple IDs, at least one of which is always critical - corresponding to a minimal subset of the generators.  It should now be clear that the complete listing of critical IDs also automatically includes the complete listing of all minimal entangled sets of generators, and thus that the IDs fully characterize the entanglement properties of the $N$-qubit Pauli group.

As a final point, we have not proved the reverse statement relating entanglement to commutativity, but we conjecture that it is also true.  Specifically, if we know that a given multipartite state $|\psi\rangle$ is maximally entangled, is it then true that there must exist an entangled set of observables with $|\psi\rangle$ as a joint eigenstate?  In the case of the $N$-qubit Pauli group the answer appears to be yes, since we can construct all of the Graph stabilizer groups and see explicitly (at least up to $N=5$) that all of these contain critical IDs, but even for the Pauli group we do not have a formal proof of this point.\pagebreak

\subsubsection{}
\begin{table}[h!]
\centering
\qquad
\subfloat[][ID$4^4_2$]{
\begin{tabular}{cccc}
$Z$ & $Z$ & $Z$ & $Z$ \\
$X$ & $X$ & $X$ & $X$ \\
$Y$ & $I$ & $Z$ & $X$ \\
$I$ & $Y$ & $X$ & $Z$ \\
\end{tabular}\label{ID4_q4_1}}
\qquad
\subfloat[][ID$4^4_4$]{
\begin{tabular}{cccc}
$Z$ & $Z$ & $Z$ & $I$ \\
$X$ & $X$ & $I$ & $Z$ \\
$Y$ & $I$ & $X$ & $X$ \\
$I$ & $Y$ & $Y$ & $Y$ \\
\end{tabular}\label{ID4_q4_2}}
\qquad
\subfloat[][ID$5^4_0$]{
\begin{tabular}{cccc}
$Z$ & $Z$ & $Z$ & $Z$ \\
$Z$ & $Z$ & $X$ & $X$ \\
$X$ & $X$ & $I$ & $I$ \\
$X$ & $I$ & $Z$ & $X$ \\
$I$ & $X$ & $X$ & $Z$ \\
\end{tabular}\label{ID5_q4_1}}
\qquad
\subfloat[][ID$5^4_2$]{
\begin{tabular}{cccc}
$Z$ & $Z$ & $Z$ & $Z$ \\
$X$ & $I$ & $X$ & $I$ \\
$Y$ & $I$ & $Z$ & $X$ \\
$I$ & $X$ & $X$ & $Z$ \\
$I$ & $Y$ & $I$ & $X$ \\
\end{tabular}\label{ID5_q4_3}}
\qquad
\subfloat[][ID$5^4_2$]{
\begin{tabular}{cccc}
$Z$ & $Z$ & $Z$ & $I$ \\
$X$ & $X$ & $I$ & $Z$ \\
$Y$ & $I$ & $X$ & $X$ \\
$I$ & $Y$ & $X$ & $X$ \\
$I$ & $I$ & $Z$ & $Z$ \\
\end{tabular}\label{ID5_q4_4}}
\qquad
\subfloat[][ID$5^4_2$]{
\begin{tabular}{cccc}
$Z$ & $Z$ & $Z$ & $I$ \\
$X$ & $X$ & $Z$ & $Z$ \\
$Y$ & $Z$ & $X$ & $X$ \\
$I$ & $Z$ & $I$ & $X$ \\
$I$ & $Y$ & $X$ & $Z$ \\
\end{tabular}\label{ID5_q4_5}}
\qquad
\subfloat[][ID$5^4_2$]{
\begin{tabular}{cccc}
$Z$ & $Z$ & $Z$ & $I$ \\
$X$ & $X$ & $I$ & $Z$ \\
$Y$ & $Z$ & $X$ & $Z$ \\
$I$ & $Z$ & $Z$ & $X$ \\
$I$ & $Y$ & $X$ & $X$ \\
\end{tabular}\label{ID5_q4_6}}
\qquad
\subfloat[][ID$5^4_2$]{
\begin{tabular}{cccc}
$Z$ & $Z$ & $Z$ & $I$ \\
$Z$ & $X$ & $X$ & $Z$ \\
$Z$ & $Y$ & $X$ & $X$ \\
$X$ & $X$ & $Z$ & $Z$ \\
$Y$ & $X$ & $I$ & $X$ \\
\end{tabular}\label{ID5_q4_7}}
\qquad\qquad\qquad\qquad\qquad\qquad
\subfloat[][]{
\includegraphics[width=.75in]{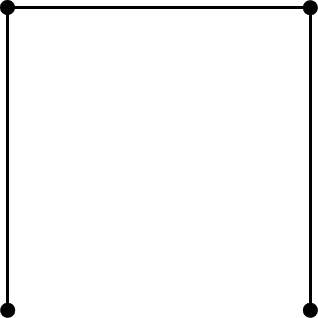}
\label{g4_1}}
\qquad
\subfloat[][]{
\includegraphics[width=.75in]{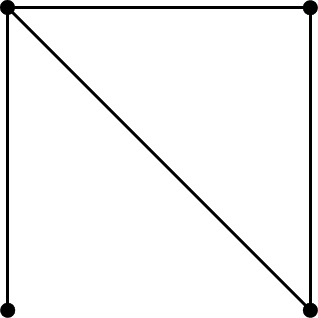}
\label{g4_3}}
\qquad
\subfloat[][]{
\includegraphics[width=.75in]{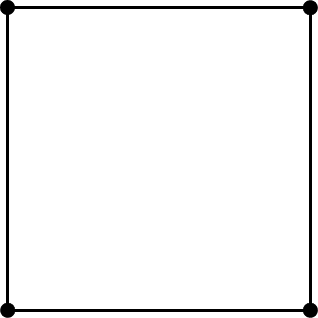}
\label{g4_4}}
\qquad
\subfloat[][]{
\includegraphics[width=.75in]{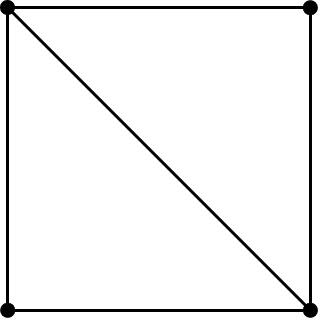}
\label{g4_5}}
\qquad
\caption[]{These 8 critical ID5s belong to the stabilizer groups of the 4-qubit Cluster states, also generated by these graphs.  The Whole ID of \subref{ID5_q4_1} is the only NKS set for 4 qubits with only 1 ID.  The two $ID4^4$s of \subref{ID4_q4_1} and \subref{ID4_q4_2} are the smallest examples of critical ID$M^N$s with $M<N+1$.}\label{q4IDs_8}
\end{table}
\begin{figure}[h!]
\centering
\qquad
\subfloat[][]{
\includegraphics[width=1.3in]{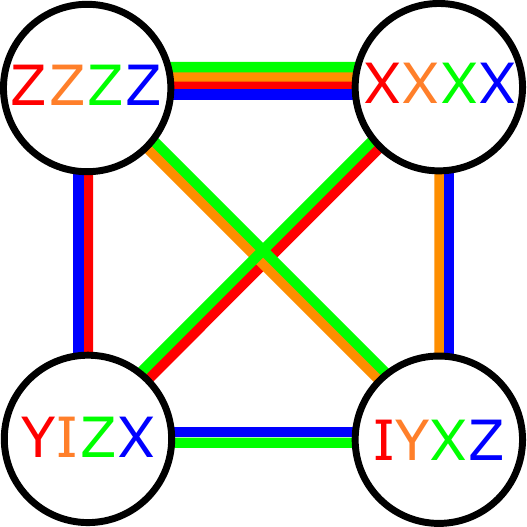}
\label{IDG4_1}}
\qquad
\subfloat[][]{
\includegraphics[width=1.3in]{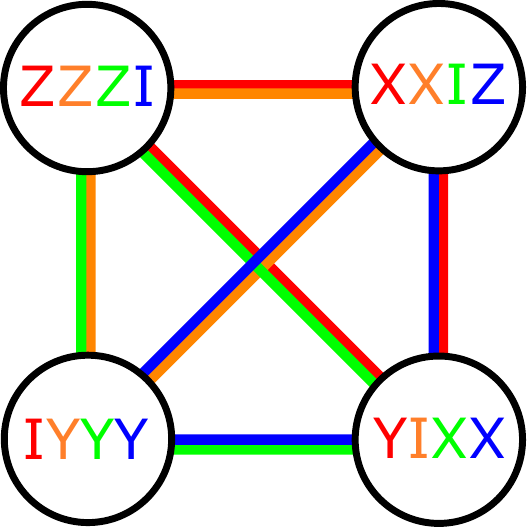}
\label{IDG4_2}
}
\qquad
\subfloat[][]{
\includegraphics[width=1.4in]{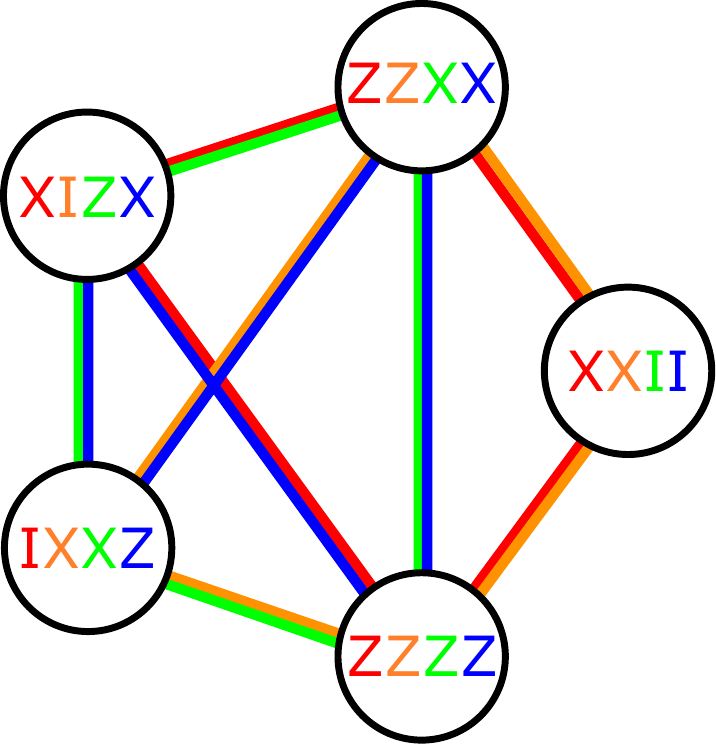}
\label{IDG4_3}
}
\qquad
\subfloat[][]{
\includegraphics[width=1.4in]{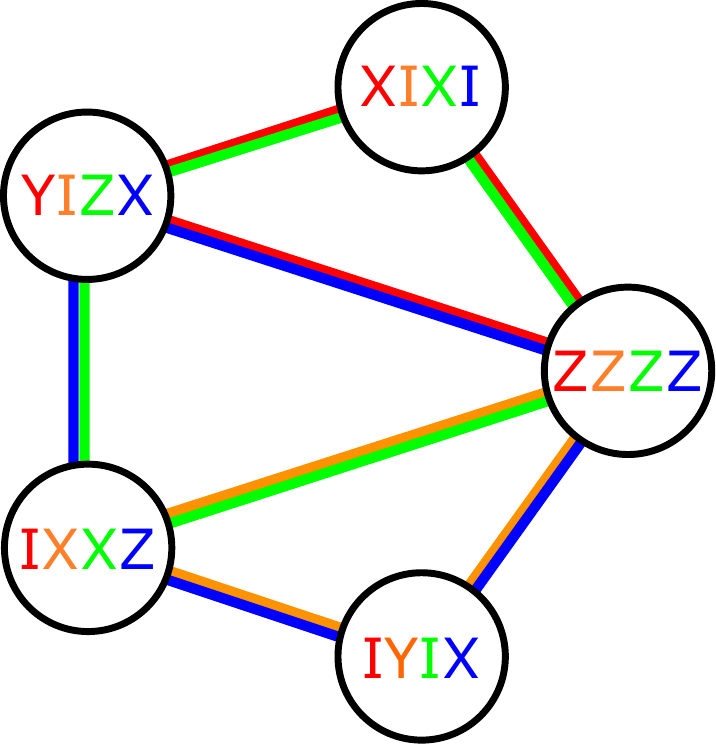}
\label{IDG4_4}}
\qquad
\subfloat[][]{
\includegraphics[width=1.4in]{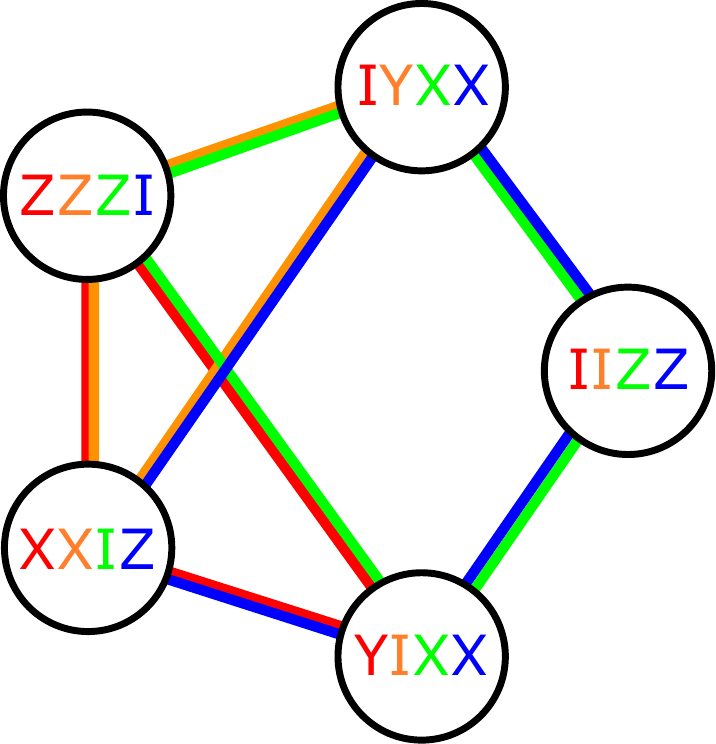}
\label{IDG4_5}
}
\qquad
\subfloat[][]{
\includegraphics[width=1.4in]{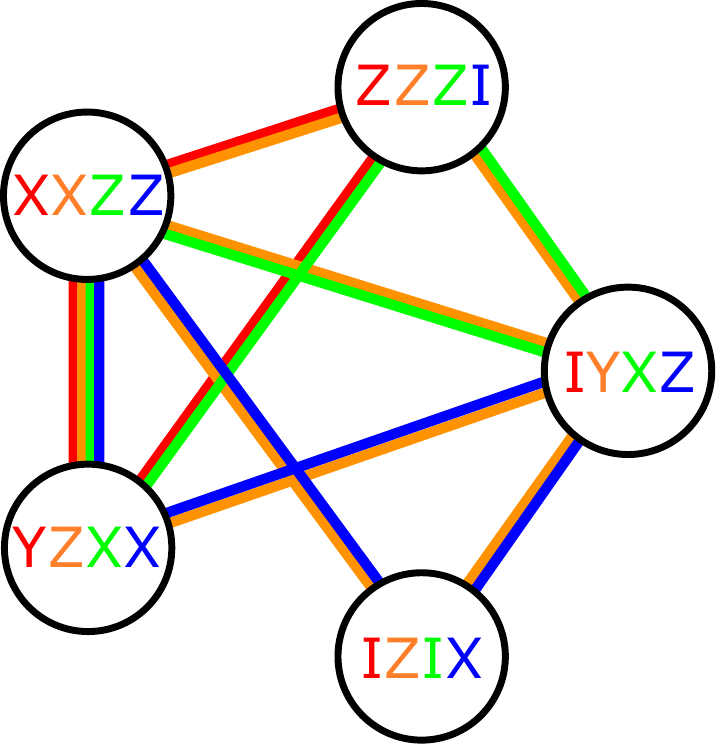}
\label{IDG4_6}
}
\qquad
\subfloat[][]{
\includegraphics[width=1.4in]{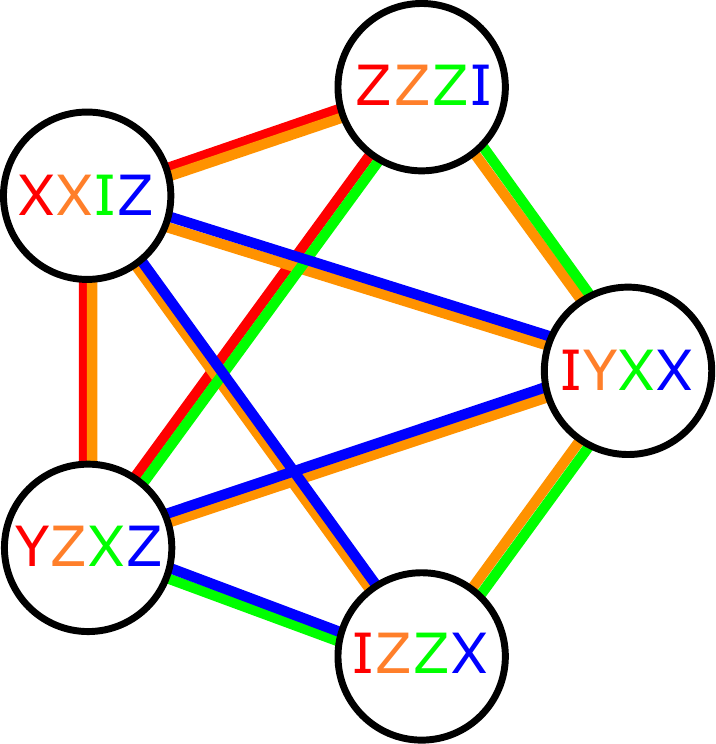}
\label{IDG4_7}}
\qquad
\subfloat[][]{
\includegraphics[width=1.4in]{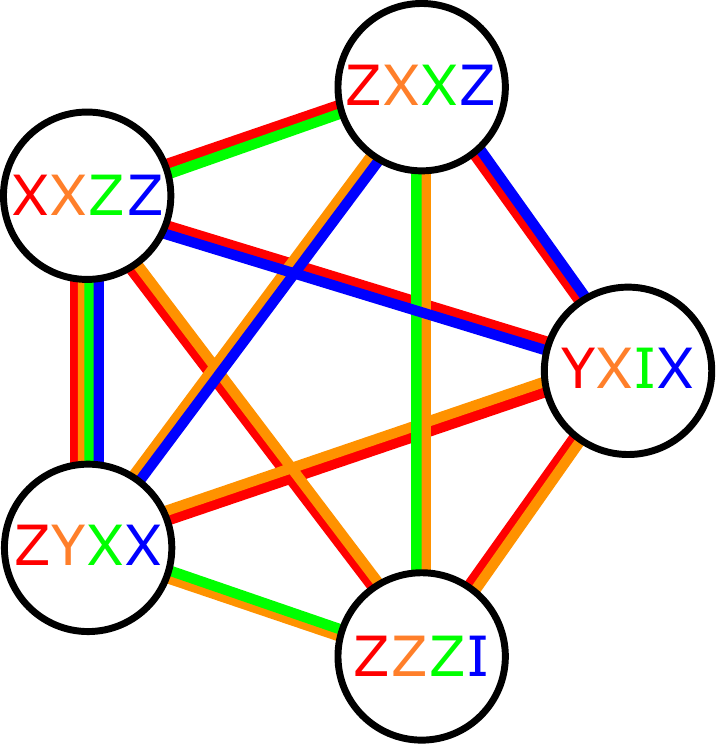}
\label{IDG4_8}
}
\qquad
\subfloat[][]{
\includegraphics[width=1.4in]{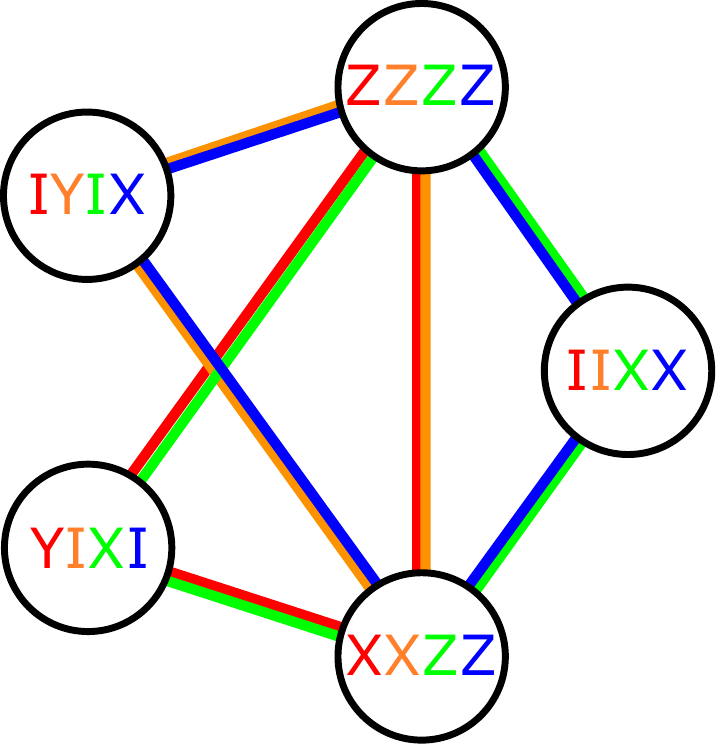}
\label{IDG4_9}
}
\qquad
\caption[]{(Color online) The 9 nonisomorphic color-hypergraphs for all representative critical IDs with 4 qubits.  The ID of \subref{IDG4_9} belongs to the 4-qubit GHZ entanglement class, while the others belong to the 4-qubit Cluster entanglement class.}\label{IDGraphs2}
\end{figure}
\begin{table}[h!]
\centering
\subfloat[][ID$5^7_2$]{
\begin{tabular}{ccccccc}
${Z}$ & $Z$ & $Z$ & $Z$ & $Z$ & $I$ & $I$ \\
${X}$ & $X$ & $Z$ & $X$ & $X$ & $Z$ & $Z$ \\
$Y$ & $I$ & $X$ & $Z$ & $Z$ & $X$ & $X$ \\
$I$ & $Y$ & $I$ & $X$ & $I$ & $Z$ & $X $ \\
$I$ & $I$ & $X$ & $I$ & $X$ & $X$ & $Z $ \\
\end{tabular} \label{M5N7}}
\qquad
\subfloat[][ID$6^{11}_2$]{
\begin{tabular}{ccccccccccc}
$ I$ & $I$ & $Z$ & $Z$ & $Z$ & $Z$ & $Z$ & $Z$ & $Z$ & $Z$ & $Z  $ \\
$ {Z}$ & $I$ & $Z$ & $Z$ & $Z$ & $X$ & $X$ & $I$ & $X$ & $X$ & $I $ \\
$ I$ & $Z$ & $X$ & $X$ & $I$ & $Z$ & $Z$ & $Z$ & $I$ & $I$ & $I $ \\
$ {X}$ & $I$ & $X$ & $I$ & $X$ & $X$ & $I$ & $X$ & $Z$ & $X$ & $X $ \\
$ I$ & $X$ & $I$ & $X$ & $I$ & $I$ & $X$ & $I$ & $I$ & $Z$ & $Z $ \\
$ Y$ & $Y$ & $I$ & $I$ & $X$ & $I$ & $I$ & $X$ & $X$ & $I$ & $X $ \\
\end{tabular} \label{M6N11}}
\qquad
\subfloat[][ID$7^{16}_{16}$]{
\begin{tabular}{cccccccccccccccc}
${Z}$ & $Z$ & $Z$ & $Z$ & $Z$ & $Z$ & $Z$ & $Z$ & $I$ & $I$ & $I$ & $I$ & $I$ & $I$ & $I$ & $I $ \\
${X}$ & $X$ & $X$ & $X$ & $I$ & $I$ & $I$ & $I$ & $Z$ & $Z$ & $Z$ & $Z$ & $I$ & $I$ & $I$ & $I $ \\
$Y$ & $I$ & $I$ & $I$ & $X$ & $X$ & $X$ & $I$ & $X$ & $I$ & $I$ & $I$ & $Z$ & $Z$ & $I$ & $I $ \\
$I$ & $Y$ & $I$ & $I$ & $Y$ & $I$ & $I$ & $I$ & $I$ & $X$ & $X$ & $X$ & $X$ & $I$ & $Z$ & $I $ \\
$I$ & $I$ & $I$ & $I$ & $I$ & $Y$ & $I$ & $X$ & $Y$ & $Y$ & $I$ & $I$ & $I$ & $I$ & $X$ & $Z $ \\
$I$ & $I$ & $Y$ & $I$ & $I$ & $I$ & $I$ & $Y$ & $I$ & $I$ & $Y$ & $I$ & $Y$ & $X$ & $I$ & $X $ \\
$I$ & $I$ & $I$ & $Y$ & $I$ & $I$ & $Y$ & $I$ & $I$ & $I$ & $I$ & $Y$ & $I$ & $Y$ & $Y$ & $Y $ \\
\end{tabular}\label{Kite7_16}}
\qquad
\caption[Maximally Compact IDs]{Examples of critical Partial IDs with the largest known $N$ for a given value of $M$.}\label{SmallIDs}
\end{table}
\begin{table}[h!]
\centering
\qquad
\subfloat[][]{
\begin{tabular}{cc}
O & O \\
O & O \\
\end{tabular}\label{CKS2}}
\qquad
\subfloat[][]{
\begin{tabular}{ccc}
O & O & I \\
I & O & O \\
O & I & O \\
\end{tabular}\label{CKS3}}
\qquad
\subfloat[][]{
\begin{tabular}{cccc}
O & O & O & O \\
O & O & O & O \\
\end{tabular}\label{CKS4_2}}
\qquad
\subfloat[][]{
\begin{tabular}{cccc}
O & O & O & O \\
O & O & I & I \\
I & I & O & O \\
\end{tabular}\label{CKS4_3}}
\qquad
\subfloat[][]{
\begin{tabular}{cccc}
O & O & O & O \\
O & O & I & I \\
O & I & O & I \\
O & I & I & O \\
\end{tabular}\label{CKS4_4_1}}
\qquad
\subfloat[][]{
\begin{tabular}{cccc}
O & O & I & I \\
I & O & O & I \\
I & I & O & O \\
O & I & I & O \\
\end{tabular}\label{CKS4_4_2}}
\qquad
\subfloat[][]{
\begin{tabular}{ccccc}
O & O & O & O & I \\
O & O & O & I & O \\
I & I & I & O & O \\
\end{tabular}\label{CKS5_3}}
\qquad
\subfloat[][]{
\begin{tabular}{ccccc}
O & O & O & O & I \\
O & O & O & I & O \\
O & O & I & O & O \\
O & O & I & I & I \\
\end{tabular}\label{CKS5_4_1}}
\qquad
\subfloat[][]{
\begin{tabular}{ccccc}
O & O & O & O & I \\
O & O & O & I & O \\
O & I & I & O & I \\
O & I & I & I & O \\
\end{tabular}\label{CKS5_4_2}}
\qquad
\subfloat[][]{
\begin{tabular}{ccccc}
O & O & O & O & I \\
O & O & I & I & I \\
I & I & O & I & O \\
I & I & I & O & O \\
\end{tabular}\label{CKS5_4_3}}
\qquad
\subfloat[][]{
\begin{tabular}{ccccc}
O & O & O & O & I \\
O & O & O & I & O \\
O & O & I & O & O \\
O & I & O & O & O \\
I & O & O & O & O \\
\end{tabular}\label{CKS5_5_1}}
\qquad
\subfloat[][]{
\begin{tabular}{ccccc}
O & O & O & O & I \\
O & O & O & I & O \\
O & O & I & O & O \\
O & I & O & I & I \\
I & O & O & I & I \\
\end{tabular}\label{CKS5_5_2}}
\qquad
\subfloat[][]{
\begin{tabular}{ccccc}
O & O & O & O & I \\
O & O & O & I & O \\
O & O & I & I & I \\
O & I & I & O & I \\
I & O & I & I & O \\
\end{tabular}\label{CKS5_5_3}}
\qquad
\subfloat[][]{
\begin{tabular}{ccccc}
O & O & O & O & I \\
O & O & I & I & I \\
O & I & O & I & I \\
O & I & I & I & O \\
I & I & I & O & O \\
\end{tabular}\label{CKS5_5_4}}
\qquad
\subfloat[][]{
\begin{tabular}{ccccc}
O & O & O & O & I \\
O & I & I & I & O \\
I & O & I & I & O \\
I & I & O & I & O \\
I & I & I & O & O \\
\end{tabular}\label{CKS5_5_5}}
\qquad
\subfloat[][]{
\begin{tabular}{ccccc}
O & O & I & I & I \\
I & O & O & I & I \\
I & I & O & O & I \\
I & I & I & O & O \\
O & I & I & I & O \\
\end{tabular}\label{CKS5_5_6}}
\qquad
\caption{The 16 Critical NKS Structures (CNSs) for $O=2,3,4,5$ qubits.}\label{CNSs}
\end{table}

\end{document}